\documentclass[8.5pt,twoside,twocolumn]{article}
\usepackage[super,sort&compress,comma]{natbib} 
\usepackage{mhchem}
\usepackage{times,mathptmx}
\usepackage{sectsty}
\usepackage{balance} 

\usepackage{graphicx} 
\usepackage{lastpage}
\usepackage[format=plain,justification=raggedright,singlelinecheck=false,font=small,labelfont=bf,labelsep=space]{caption} 
\usepackage{fancyhdr}
\usepackage{amssymb,amsmath}

\usepackage{amsmath}		
\usepackage{subfigure}		
\usepackage{color}
\usepackage{xcolor}		
\usepackage{bm}			
\usepackage{mathrsfs}
\usepackage{cases}

\newcommand{\mr}{\mathrm}

\newcommand{\mc}{\mathcal}

\newcommand{\md}{\mathrm{d}}
\newcommand{\vD}{\varDelta}

\begin{document}

\thispagestyle{plain}
\fancypagestyle{plain}{
\fancyhead[L]{\includegraphics[height=8pt]{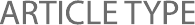}}
\fancyhead[C]{\hspace{-1cm}\includegraphics[height=20pt]{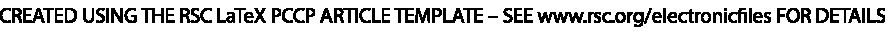}}
\fancyhead[R]{\includegraphics[height=10pt]{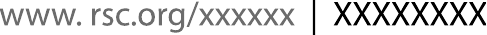}\vspace{-0.2cm}}
\renewcommand{\headrulewidth}{1pt}}
\renewcommand{\thefootnote}{\fnsymbol{footnote}}
\renewcommand\footnoterule{\vspace*{1pt}%
\hrule width 3.4in height 0.4pt \vspace*{5pt}} 
\setcounter{secnumdepth}{5}

\makeatletter
\DeclareRobustCommand\onlinecite{\@onlinecite}
\def\@onlinecite#1{\begingroup\let\@cite\NAT@citenum\citealp{#1}\endgroup}
\makeatother

\makeatletter 
\def\subsubsection{\@startsection{subsubsection}{3}{10pt}{-1.25ex plus -1ex minus -.1ex}{0ex plus 0ex}{\normalsize\bf}} 
\def\paragraph{\@startsection{paragraph}{4}{10pt}{-1.25ex plus -1ex minus -.1ex}{0ex plus 0ex}{\normalsize\textit}} 
\renewcommand\@biblabel[1]{#1}            
\renewcommand\@makefntext[1]%
{\noindent\makebox[0pt][r]{\@thefnmark\,}#1}
\makeatother 
\renewcommand{\figurename}{\small{Fig.}~}
\sectionfont{\large}
\subsectionfont{\normalsize} 

\fancyfoot{}
\fancyfoot[LO,RE]{\vspace{-7pt}\includegraphics[height=9pt]{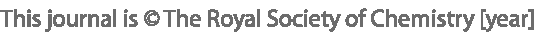}}
\fancyfoot[CO]{\vspace{-7.2pt}\hspace{12.2cm}\includegraphics{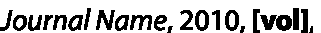}}
\fancyfoot[CE]{\vspace{-7.5pt}\hspace{-13.5cm}\includegraphics{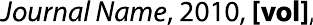}}
\fancyfoot[RO]{\footnotesize{\sffamily{1--\pageref{LastPage} ~\textbar  \hspace{2pt}\thepage}}}
\fancyfoot[LE]{\footnotesize{\sffamily{\thepage~\textbar\hspace{3.45cm} 1--\pageref{LastPage}}}}
\fancyhead{}
\renewcommand{\headrulewidth}{1pt} 
\renewcommand{\footrulewidth}{1pt}
\setlength{\arrayrulewidth}{1pt}
\setlength{\columnsep}{6.5mm}
\setlength\bibsep{1pt}

\twocolumn[
  \begin{@twocolumnfalse}
\noindent\LARGE{\textbf{Monte-Carlo Simulations of Spin-Crossover Phenomena Based on a Vibronic Ising-like Model with Realistic Parameters}}
\vspace{0.6cm}

\noindent\large{\textbf{Hongzhou Ye,$^{\ddag}$ Chong Sun,\textit{$^{\ddag{}}$} and
Hong Jiang$^{\ast}$}}\vspace{0.5cm}

\noindent\textit{\small{\textbf{Received Xth XXXXXXXXXX 20XX, Accepted Xth XXXXXXXXX 20XX\newline
First published on the web Xth XXXXXXXXXX 200X}}}

\noindent \textbf{\small{DOI: 10.1039/b000000x}}
\vspace{0.6cm}

\noindent \normalsize{Materials with spin-crossover (SCO) properties hold great potentials in information storage and therefore have received a lot of concerns in the recent decades. The hysteresis phenomena accompanying SCO is attributed to the intermolecular cooperativity whose underlying mechanism may have a vibronic origin. In this work, a new vibronic Ising-like model in which the elastic coupling between SCO centers is included by considering harmonic stretching and bending (SAB) interactions is proposed and solved by Monte Carlo simulations. The key parameters in the new model, $k_1$ and $k_2$, corresponding to the elastic constant of the stretching and bending mode, respectively, can be directly related to the macroscopic bulk and shear modulus of the material in study, which can be readily estimated either based on experimental measurements or first-principles calculations. The convergence issue in the MC simulations of the thermal hysteresis has been carefully checked, and it was found that the stable hysteresis loop can be more readily obtained when using the SAB model compared to that using the Wajnflasz-Pick model. Using realistic parameters estimated based on first-principles calculations of a specific polymeric coordination SCO compound, \ce{[Fe(pz)Pt(CN)4].2H2O}, temperature-induced hysteresis and pressure effects on SCO phenomena are simulated successfully. }
\vspace{0.5cm}
 \end{@twocolumnfalse}
  ]

\section{Introduction} 
\footnotetext{\textit{Beijing National Laboratory for Molecular Sciences, College of Chemistry and Molecular Engineering, Peking University, 100871 Beijing, China.}}


\footnotetext{\ddag~These authors contributed equally to this work.}

		Many transition metal complexes with d$^4$ to d$^7$ electronic configurations in octahedral coordination environments are able to undergo transition between low-spin (LS) and high-spin (HS) states, often termed as spin-crossover (SCO), under perturbations such as variation of temperature or pressure, light irradiation, applied electric or magnetic field. \cite{Kahn98, Boca03, Bousseksou11CSR} It is ``one of the most spectacular examples of molecular bistability" as O. Kahn remarked, and may function as ``active elements in memory devices".  \cite{Kahn98} Because of their promising applications in information storage as well as their intertest in fundamental study of phase transitions, SCO systems have attracted tremendous intertest both experimentally and theoretically in past decades. \cite{Guetlich04Book, Guetlich94ACIE, Kahn98, Boca03, Halcrow11CSR, Tao12CSR, Pavlik13EJIC, Paulsen13EJIC} The molecular origin of SCO phenomena can be qualitatively understood based on ligand field theory, \cite{Kahn98} and modern electronic structure theory have made great progress on quantitative prediction of SCO properties, including, in particular, the energy splitting between LS and HS states $\varDelta_{\rm HL}$. \cite{Paulsen13EJIC} 
		
		One of the most spectacular features of SCO systems is the cooperativity in the SCO processes. A significant volume change usually accompanies the LS-HS transition, in which one or two electrons transfer from the non-bonding $t_{2g}$ orbitals to the antibonding $e_g$ orbitals. Wide hysteresis loops may be obtained as a result of strong coupling between intramolecular SCO and intermolecular interactions.\cite{Real03CCR, Vreug90, Kahn98} Thus the polymeric coordination SCO materials which have a stronger intermolecular interaction are promising to give wider hysteresis loops. 2D and 3D \ce{Fe(II)}-centered Hofmann-like frames have been synthesized and characterized in many laboratories,\cite{Kitazawa96, Molnar04, Hosoya03, Niel01} some of which have shown a relatively wide hysteresis loop around room temperature.    
		
		Simulations of SCO systems based on Monte-Carlo methods have received widespread concerns, and several theoretical models have been proposed (See, e.g. Refs. \onlinecite{Pavlik13EJIC, Bousseksou11CSR} for a comprehensive review). Ising-like models, e.g., the Wajnflasz-Pick (WP) model with a site-independent and somewhat phenomenological interaction parameter $J^{\mr{WP}}$, have been developed prosperously and explained various aspects of SCO behaviours. \cite{Wajnflasz71, Kahn95, Nishino03, Bous92} Many of them have been generalized to model nanosized SCO compounds \cite{Bousseksou11CSR, Mikolasek14} by adjusting the boundary conditions used in simulations. \cite{Enach14, Mikolasek14} However, most simulations with the WP model have to make a compromise between the acceptable computation time and stable results,\cite{A.Atitoaie12,Enach14,Enach11} since hysteresis loops obtained in this way often shrink when slower sweeping rates (i.e. number of simulation steps for each temperature point) are employed. On the other hand, mechanoelastic models (or atom-phonon model,\cite{J.A.Nasser01,K.Boukheddaden07} vibronic Ising-like model) such as ``ball and spring" model, attribute the phenomenological interaction parameter $J^{\mr{WP}}$ in the WP model to the elastic interaction arising from molecular volume change during SCO. \cite{Konishi08, Enach11} However, parameters of interaction strengths in those models are determined from simulations results rather than experimental work or \textit{ab initio} calculations, making them not enough to describe realistic models.
		
		This paper is organized as follows. In the first part, we briefly review existing theoretical models and then propose the ``stretching and bending" (SAB) model whose interaction parameters can be determined either experimentally or \textit{ab initio}. Both the elastic (up to the three-body level) and the vibronic interactions are shown to be covered in this model. In the second part, details regarding simulation strategies and parameter values used in this paper are stated. In the third part, we first discuss the determination of key parameters $k_1$ and $k_2$ from density functional theory (DFT) calculations of a specific ferrous Hoffman-like compound, \ce{[Fe(pz)Pt(CN)4].2H2O}. Then we show that simulations with the SAB model are able to reach a stable result. Finally we demonstrate that it is able to simulate both temperature-induced hysteresis and pressure effects on SCO phenomena.

	\section{Models}
	    \subsection{Ising-like models}
		Consider a solid material constituted of $N$ SCO molecules, each of which has two spin states: HS and LS, labelled with the fictitious spin variable $s=\pm{}1$, respectively. The Ising-like Hamiltonian can be generally writen as  
		\begin{equation} \label{eq:exact Ising-model}
			\mathcal{H}(\{s_i\},T) =
				\frac{1}{2}h(T)\sum_{i=1}s_i
				-\frac{1}{2}\sum_{\langle{}i,j\rangle}J(s_i,s_j)s_i{}s_j.
		\end{equation}
		The first term in eqn (\ref{eq:exact Ising-model}) is given by 
		\begin{equation}
			h(T) = 
				\vD(T)-k_{\mr{B}} T \ln{}g(T), 
		\end{equation}				
		$\vD(T)$ is the intramolecular energy difference accounting for both the electronic HS-LS splitting $\vD_{\rm HL}$ and the temperature-dependent vibrational energy difference $\Delta E_{\rm vib}(T)$ between HS and LS states, and $g(T)$ is the ratio between the effective degeneracy in the LS and HS state, $g(T) = g_{\mr{HS}}(T)/g_{\mr{LS}}(T)$. \cite{Miyashita05} The second term in eqn (\ref{eq:exact Ising-model}) describes the interactions between SCO centers, which in most cases consider only the nearest neighbouring interaction. By comparing eqn (\ref{eq:exact Ising-model}) to the phenomenological equation by Slichter and Drickamer,\cite{Slichter72} one can obtain \cite{Pavlik13EJIC} the relation
		\begin{equation}	\label{eq:macro-micro}
			\Delta{}H(T)= N_{\mr{A}}\vD(T),\quad
			\Delta{}S(T)= R\ln{}g(T).
		\end{equation}
		In principle, eqn (\ref{eq:macro-micro}) enables us to determine the values of $\vD(T)$ and $g(T)$ from either experimental data or \textit{ab initio} calculations. However, only $\Delta{}H(T_{1/2})$ and $\Delta{}S(T_{1/2})$ evaluated at the transition temperature $T_{1/2}$ can be measured experimentally. On the other hand, it is also not trivial to determine temperature-dependent enthalpy and entropy from theoretical calculations, which require the electronic energy difference at zero temperature ($\Delta E_{\rm HL}$) and the full vibrational (phonon) spectrum. 
		 
		
	     \subsection{The WP Model}		
		The WP model\cite{Wajnflasz71} is an approximation of the general Ising-like model (\ref{eq:exact Ising-model}), in which experimentally determinable $\vD(T_{1/2})$, $g(T_{1/2})$ and a site-independent interaction parameter $J^{\mr{WP}}$ are used:
		\begin{equation}	\label{eq:prototype Ising model}
			\mathcal{H}^{\mr{WP}}(\{s_i\},T) = 
				\frac{1}{2}h^{\mr{WP}}(T)\sum_{i=1}s_i
				-\frac{J^{\mr{WP}}}{2}\sum_{\langle{}i,j\rangle}s_i{}s_j,
		\end{equation}		
		where $h^{\mr{WP}}(T)=\vD(T_{1/2})-k_{\mr{B}}T\ln{}g(T_{1/2})$. In this model the summation of interaction terms is restricted to the nearest pair $\langle{}i,j\rangle$. This approximation makes a mean-field analysis (MFA) possible, which gives a non-trivial criterion for phase transition accompanied with hysteresis\cite{Boukheddaden00a, Varret03, Nishino03}
		\begin{equation}	\label{eq:judgement_hysteresis}
			J^{\mr{WP}}\geq{}J_{\mr{thresh}}=2\vD/(z\ln{}g),
		\end{equation}
		where $z$ is the coordination number under MFA. A generalization to the exact case (i.e. get rid of MFA) is straightforward, $J_{\mr{thresh}}=2\vD/(B\ln{}g)$, where $B$ is the coefficient in the expression of Ising model's critical temperature, $T_{\mr{c}}=BJ/k_{\mr{B}}$ (e.g. $B=2.269185$ for a 2D square lattice).
		
		\subsection{The ``Ball and Spring" Model}
		If lattice vibrations are added to the Ising-like model through harmonic oscillators between molecules, we have the vibronic Ising-like model. One example is the ``ball and spring" (BAS) model. \cite{Enach11, Konishi08} In this model, degrees of freedom of molecular positions $\{\bm{r}_i\}$ are included and the volume difference between HS and LS states is treated explicitly. Interactions between the nearest ($\langle{}i,j\rangle$) and the second-nearest pair ($\langle\!\langle{}i,j\rangle\!\rangle$) are assumed to be harmonic:
		\begin{equation}	\label{eq:SHO Ising model}
		\begin{split}
			&\mathcal{H}^{\mr{BAS}}(\{x_i\},T)=
				\frac{1}{2}h^{\mr{WP}}(T)\sum_{i=1}s_i-\mathcal{H}_1^{\mr{BAS}}-\mathcal{H}_2^{\mr{BAS}},		\\
			&\mathcal{H}^{\mr{BAS}}_{1}(\{x_i\})=\frac{1}{2}\sum_{\langle{}i,j\rangle}
				\frac{k_1}{2}\big[r_{ij}-(R_{i}+R_{j})\big]^2,							\\
			&\mathcal{H}^{\mr{BAS}}_{2}(\{x_i\})=\frac{1}{2}\sum_{\langle\!\langle{}i,j\rangle\!\rangle}
				\frac{k_2}{2}\big[r_{ij}-\sqrt{2}(R_{i}+R_{j})\big]^2.
		\end{split}
		\end{equation}
		where $x=(\bm{r},s)$ is the four-component dynamic variable, $R_{i}=[(1+s_i)R_{\mr{HS}}+(1-s_i)R_{\mr{LS}}]/2$ is the radius of molecule at site $i$ with spin state $s_i$, and $r_{ij}=|\bm{r}_i-\bm{r}_j|$ is the distance between site $i$ and $j$. Both experimental data and \textit{ab initio} study support a larger-than-unity ratio $R_{\mr{HS}}/R_{\mr{LS}}$. By expanding $\mc{H}_1$ and $\mc{H}_2$ one can show that they are actually equivalent to a site-dependent interaction term $J(x_i,x_j)$, which is just a direct generalization of $J(s_i,s_j)s_i{}s_j$ in eqn (\ref{eq:exact Ising-model}). We take $\mc{H}_1^{\mr{BAS}}$ as an example:
		\begin{equation}	\label{eq:expansion_H_1}
			\begin{split}
				\frac{k_1}{2}[r_{ij}-&(R_i+R_j)]^2=\frac{k_1}{4}
					(R_{\mr{HS}}-R_{\mr{LS}})^{2}s_i{}s_j+									\\
				&\frac{k_1}{2}(R_{\mr{HS}}+R_{\mr{LS}}-r_{ij})
					(R_{\mr{HS}}-R_{\mr{LS}})(s_i+s_j)+										\\
				&\frac{k_1}{2}(R_{\mr{HS}}+R_{\mr{LS}}-r_{ij})^2+
					\frac{k_1}{4}(R_{\mr{HS}}-R_{\mr{LS}})^2.
			\end{split}	
		\end{equation}		 
		The first and third term in eqn (\ref{eq:expansion_H_1}) depend merely on $\{s_i\}$ or $\{\bm{r}_i\}$ and characterize interactions araising from either spins or lattice distortions respectively, while the second one containing crossing term $r_{ij}s_i$ (or $r_{ij}s_j$) describes the spin-vibration coupling. The last term however, is a constant and can be removed. More importantly, if compared with the WP model (\ref{eq:prototype Ising model}), one could relate the first term in eqn (\ref{eq:expansion_H_1}) to $J^{\mr{WP}}$ as
		\begin{equation}	\label{eq:WP_vs_BAS}
			J^{\mr{WP}}\sim{}\frac{k_1}{4}(R_{\mr{HS}}-R_{\mr{LS}})^2.
		\end{equation}
		Relation (\ref{eq:WP_vs_BAS}) makes it possible to compare these two models directly: apart from spin-spin interactions, the BAS model includes both vibrational and vibronic interactions as well. 
		
		 Model (\ref{eq:SHO Ising model}) has been shown to successfully describe both temperature and pressure-induced hysteresis.\cite{Konishi08} $\mathcal{H}_2$ may or may not be included, depending on the particular lattice structure. For example, when a simple cubic lattice is considered, it must be included in avoid of a structural deformation,\cite{Konishi08} while it is not requisite for a hexagonal one. \cite{Enach11} 

		\subsection{The ``Stretching and Bending" Model}
\begin{figure}[!tp]
			\centering
			\includegraphics[width=0.30\textwidth]{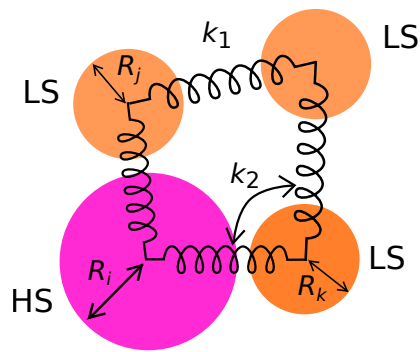}
			\caption{A schematic illustration of the SAB model. $k_1$ and $k_2$ are Hooke's coefficients for the stretching and bending springs respectively. Two states, i.e. HS and LS, have different volumes, thus leading to a lattice distortion and then elastic interactions when SCO happens.}
			\label{fig:SAB_model}
		\end{figure}

		Motivated by the concept of bond angles in chemistry, we propose that a better way to maintain the structure during simulation is to assume a harmonic potential on the angle $\theta$ between two pairs of molecules sharing one common vertex. We name it as ``stretching and bending" (SAB) model (FIG. \ref{fig:SAB_model}) for reasons we would show below. The Hamiltonian is almost the same as model (\ref{eq:SHO Ising model}) except for $\mathcal{H}_2^{\mr{BAS}}$ being replaced with a bending oscillator:
		\begin{equation}	\label{eq:polymer SHO model}
			\mathcal{H}_2^{\mr{SAB}}(\{\bm{r}_i\})=\frac{1}{2}\sum_{\langle{}i;j,k\rangle}
				\frac{k_2}{2}(\theta_{ijk}-\theta_0)^2,
		\end{equation}				
		where $\langle{}i;j,k\rangle{}$ means that site $j$ and $k$ form a non-linear nearest set with respect to site $i$ (FIG. \ref{fig:SAB_model}), and $\theta_{ijk}$ is the corresponding angle, in radius. $\theta_0$ is the equilibrium angle, whose value depends on the specific lattice (e.g. $\pi/2$ for cubic system, $\pi/6$ for hexagonal system and etc.). In contrast to the BAS model, $\mc{H}_2^{\mr{SAB}}$ depends merely on the spatial coordinates $\{\bm{r}_i\}$, which can be seen from the explicit expression of $\theta_{ijk}$. The cosine theorem gives
		\begin{equation}	\label{eq:theta_ijk}
			\theta_{ijk}=\arccos\bigg(\frac{r_{ij}^2+r_{ik}^2-r_{jk}^2}{2r_{ij}r_{ik}}\bigg).
		\end{equation}		 
		Thus interactions arising from spins or spin-vibration coupling are absent in the bending oscillator. As a compensation, a three-body interaction between $i$, $j$ and $k$ is included.
				
		Now, both $k_1$ and $k_2$ are of definite physical significance and can be related to the macroscopic bulk modulus $K$ and shear modulus $G$ respectively
		\begin{equation}	\label{eq:modulus}
			K=\frac{k_1}{3v_0^{1/3}},
			\qquad{}G=\frac{4k_2}{v_0},
		\end{equation}
		where $v_0$ is the equilibrium unit cell volume (for derivations see Appendix). Although those data of modulus are not available in laboratories now, we can easily obtain them from the equation of state (EOS) calculated by density functional theory (DFT).

	\section{Simulation Details}
		In this section we give some details on how the MC simulations are performed in practice. In this work, we consider both the WP model and the SAB model. For the WP model, we employ the canonical ($N$, $V$, $T$)-ensemble and consider a simple square lattice of $N=n^2$ sites with periodic boundary conditions. At each step in a MC simulation,  we randomly choose one site, say $i$, flip its spin $s_i$ to $-s_i$, and then accept or reject the new state according to the Metropolis probability 
		\begin{equation}	\label{eq:Metropolis_Probability}
			P(s_i\rightarrow{}-s_i)=\mr{min}[1,\exp(-\Delta{}\mc{H}/k_{\mr{B}}T)],
		\end{equation}		 	
		where $\Delta{}\mc{H}$ according to (\ref{eq:prototype Ising model}) is
		\begin{equation}
			\Delta{}\mc{H}=-h^{\mr{WP}}(T)s_i+J^{\mr{WP}}s_i\sum_{\langle{}j\rangle}s_j.
		\end{equation}
		Since larger simulation cells usually requires longer MC simulation time, we use the Monte-Carlo step (MCS) to measure the total number of simulation steps. One MCS is equal to $N$-simulation steps such that in one MCS all sites in the simulation cell are checked once on average. \cite{Enach14}
	
		For the SAB model, we consider a 3D simple cubic lattice with size $N=n^3$ and length $L=2 n R$ at each side, with $R=R_{\mr{LS}}$ or $R_{\mr{HS}}$, corresponding, respectively, to a pure LS or HS initial state. We perform the MC simulation in the Gibbs ($N$, $p$, $T$)-ensemble as in Ref. \onlinecite{Konishi08}. The Metropolis probability is now calculated in terms of
		\begin{equation}	\label{eq:NPT-MC_Metropolis} 
			P(x_i\rightarrow{}x_i')=\mr{min}\big[1,\exp(-\Delta{}W/k_{\mr{B}}T)\big],
		\end{equation}
		where
		\begin{equation}	\label{eq:Delta_W}
			\Delta{}W=\mc{H}(x_i')-\mc{H}(x_i)+p(V_i'-V_i)-NT\ln\bigg(\frac{V_i'}{V_i}\bigg).
		\end{equation}
		and
		\begin{equation}
			\begin{split}
			\mc{H}(x_i)=\frac{1}{2}h^{\mr{WP}}(T)s_i&+
				\frac{1}{2}\sum_{\langle{}j\rangle}
				\frac{k_1}{2}\big[r_{ij}-(R_{i}+R_{j})\big]^2		\\
				&+\frac{1}{2}\sum_{\langle\!\langle{}j;k\rangle\!\rangle}
				\frac{k_2}{2}(\theta_{ijk}-\pi/2)^2.
			\end{split}
		\end{equation}
		 A complete MCS includes: (i) Choose randomly a candidate site $i$ with spin $s_i$ and position $\bm{r}_i$. (ii) Set its spin to $s_i'=\pm{}1$ with the probability $g:1$. (iii) Move it to a new position $\bm{r}_i'=\bm{r}_i+\delta\bm{\xi}_{i}$, where $\delta=0.005L$, and $\xi_{\alpha}\in{}[-1.0,1.0]$ is a random number for $\alpha=x,y,z$. (iv) Update the system according to (\ref{eq:NPT-MC_Metropolis}) with $\Delta{}W=\Delta{}\mc{H}$ (i.e. no volume change in this step). (v) Repeat (i) to (iv) N times. (vi) Choose a candidate length $L_i'=L_i+\gamma{}\xi$ with $\gamma=0.08nR_{\mr{LS}}$ and $\xi$ is randomly chose from $[-1.0,1.0]$. (vii) Update $L$ according to (\ref{eq:NPT-MC_Metropolis}).
		 
		 Most parameters in simulations below are expressed in terms of Kelvin. For quantities with unit of energy (i.e. $\vD$, $J$ and $k_2$), dividing them by $k_{\mr{B}}$ completes the transform. $k_1$ needs some special treatment as followed:
		\begin{equation}
			\begin{split}
			\mathcal{H}
				&\sim{}k_1[r_{ij}-(R_i+R_j)]^2		\\
				&=k_1R_{\mr{LS}}^{\phantom{\mr{LS}}2}\bigg[\frac{r_{ij}}{R_{\mr{LS}}}-
				\bigg(\frac{R_i}{R_{\mr{LS}}}+\frac{R_j}{R_{\mr{LS}}}\bigg)\bigg]^2	\\
				&=k_1'[r_{ij}'-(R_{i}'+R_{j}')]^2,
			\end{split}
		\end{equation}
		where by choosing $R_{\mr{LS}}$ as the unity of length, we have $k_1'=k_1R_{\mr{LS}}^{\phantom{\mr{LS}}2}$ possessing the unit of energy and thus can be transformed into Kelvin. In the following, we use $k_1$, $r$ and $R$ to denote these ``reduced" quantity $k_1'$, $r'$ and $R'$ for simplicity.
	
	\section{Results and Discussions}
		\subsection{\textit{Ab Initio} Determination of $k_1$ and $k_2$}
		How to determine $k_1$ and $k_2$ should be discussed before simulations. Although we have relation (\ref{eq:modulus}) in hand, no experimental results about modulus $K$ and $G$ of these systems are available up till now. As we mentioned above however, through conducting \textit{ab initio} calculations we are able to estimate the order of magnitude of them for certain systems. For example, for the 3D simple cubic system we are concerned, $k_1$ and $k_2$ are related to these two EOSs below
		\begin{equation}	\label{eq:k1_and_k2}
		E_{\mr{el}}(x) = \frac{3}{2}k_1{}v_0^{2/3}(x-1)^2,\quad{}
		E_{\mr{el}}(\gamma)=2k_2(\gamma-\pi/2)^2,		
		\end{equation}
		where $E_{\mr{el}}$ and $v_0$ are the total electronic energy and equilibrium volume of a unit cell respectively, $x=(v/v_0)^{1/3}$, and $\gamma$ is one of the lattice angles (for derivations see Appendix).			 
		\begin{figure}[!tp]
			\centering
			\includegraphics[width=0.4\textwidth]{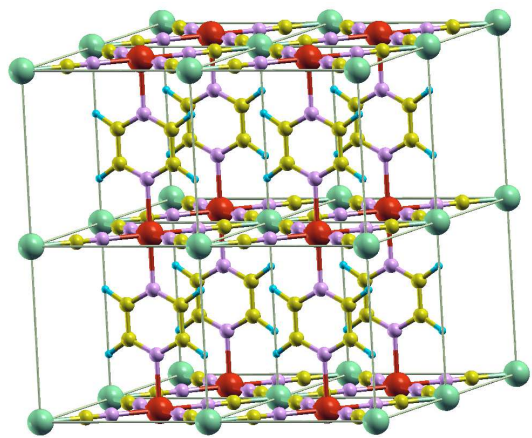}
			\caption{Schematic illustration of the crystal structure of \ce{[Fe(pz)Pt(CN)4].2H2O}. Water molecules are removed for the sake of concision.}
			\label{fig:lattice}
		\end{figure}
		\begin{figure}[btp]
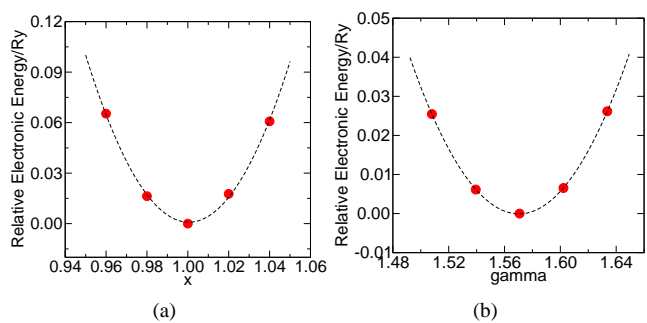

			\centering
			\subfigure[]{\includegraphics[width=0.234\textwidth]{E_vs_V.eps}}	
			\subfigure[]{\includegraphics[width=0.23\textwidth]{E_vs_cosab.eps}}
			\caption{DFT calculation results of the total electronic energy versus lattice volume (a) and angle (b) using PBEsol for \ce{[Fe(pz)Pt(CN)4].2H2O} in LS state. Dashed lines are fitting curves according to (\ref{eq:k1_and_k2}).}
			\label{fig:E_vs_cos_and_V}
		\end{figure}		
		Using (\ref{eq:k1_and_k2}), we can now estimate the approximate values for $k$'s. To realize this, we choose a ferrous Hoffman-like polymeric coordination compound with tetragonal symmetry, \ce{[Fe(pz)Pt(CN)4].2H2O} \cite{Niel01} (FIG. \ref{fig:lattice}) and conducted DFT calculations on $E_{\mr{el}}(x)$ and $E_{\mr{el}}(\gamma)$ using the PBEsol functional, \cite{Perdew08} as implemented in the Quantum ESPRESSO package.\cite{QEcode} The results are illustrated in FIG. \ref{fig:E_vs_cos_and_V}. Quadratic fittings according to (\ref{eq:k1_and_k2}) of calculated values result in good consistency and give estimated values for $k$'s in our model $k_1\approx{}1.2\times{}10^{6}\,$K and $k_2\approx{}5.5\times{}10^{5}\,$K. Thus a ratio
		\begin{equation}	\label{eq:ratio_k}
			k_1/k_2\approx{}2
		\end{equation}
		holds. We will use this ratio in all of the following simulations throughout this paper.

		\subsection{Convergence Properties}
		In simulating hysteresis phenomena, it is essential to require that the results be stable under an increase in the number of MCSs, or equivalently speaking, under slowing down the sweeping rate (MCS$/$K). This basic requisite is by no means trivial. To illustrate it, we reproduce the simulation in a work of O. Kahn's\cite{Kahn95} using the WP model with various sweeping rates. In this simulation, a $500\times{}500$ 2D square lattice is considered, with $\vD=1440\,\mr{K}$, $g=1331$ (correspondingly, $\Delta{}H=12\,\mr{kJ}\cdot{}\mr{mol}^{-1}$, $\Delta{}S=60\,\mr{J}\cdot{}\mr{mol}^{-1}\cdot{}\mr{K}^{-1}$), $T_{1/2}=200\,\mr{K}$, $J_{\mr{thresh}}=88.2\,\mr{K}$. Kahn took $J=107.9\,\mr{K}>J_{\mr{thresh}}$ so there is expected to be an abrupt phase transition with hysteresis according to (\ref{eq:judgement_hysteresis}).	
		\begin{figure}[tbp]
			\centering
			\includegraphics[width=0.40\textwidth]{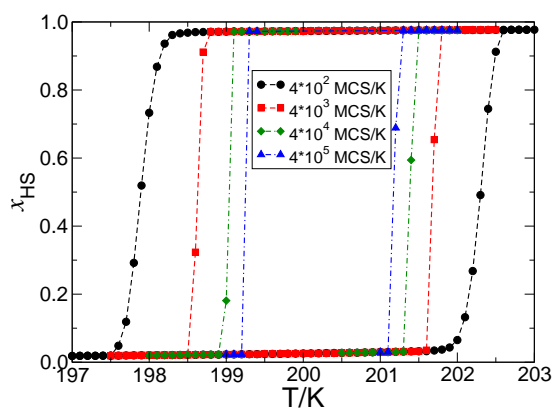}
			\caption{Reproduction of O. Kahn's simulation in ref. \onlinecite{Kahn95}, in which they claimed to obtain a stable hysteresis loop with $\Delta{}T=6\pm{}1\,\mr{K}$. Our results show that however, even with a fairly fast sweeping rate, e.g. $4\times{}10^2\,$MCS$/$K, the width of the loop is less than that value and keeps decreasing with slower sweeping rates.}
			\label{fig:Kahn_reprint}
		\end{figure}
		
		In Kahn's original paper, they claimed to obtain a stable hysteresis loop of $(6\pm1)\,$K. The results we get however, show that the loop becomes narrower when decreasing the sweeping rate (FIG. \ref{fig:Kahn_reprint}), and is less than $6\,$K even with a fairly small MCS$/$K (e.g. $4\times{}10^2$). It is clear that there is a non-ignorable convergence problem in the WP model.
				
\begin{figure}[tp]
			\centering
			\includegraphics[width=0.40\textwidth]{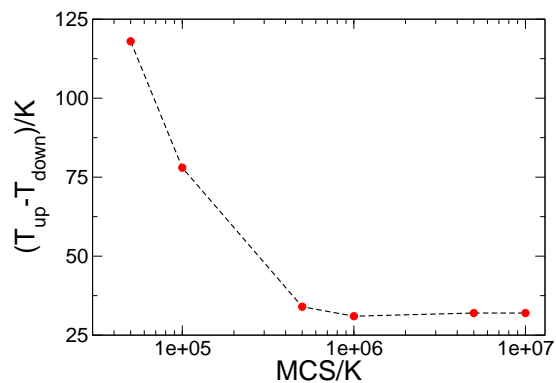}
			\caption{Hysteresis loop widths versus sweeping rates using the SAB model. $k_1=7.2\times{}10^{}4\,$K and $N=16^3$.}
			\label{fig:WP_vs_PSHO}
		\end{figure}

		A possible solution to the convergence problem is to introduce a site-dependent interaction term to replace the constant $J^{\mr{WP}}$ in the WP model. We argue that the SAB model, with a site-dependent spring interaction is an appropriate choice. To demonstrate this, we compare the widths of hysteresis loop versus sweeping rates using this model. We take typical experimentally measured $\Delta{}H=15\,\mr{kJ}\cdot{}\mr{mol}^{-1}$, $\Delta{}S=60\,\mr{J}\cdot{}\mr{mol}^{-1}\cdot{}\mr{K}^{-1}$ (correspond to $\vD=1800\,$K and $g=1360$), and the simulation is done under normal pressure (i.e. $p=1\,$atm). The results are illustrated in FIG. \ref{fig:WP_vs_PSHO}. One can clearly see that the SAB model can lead to a stable hysteresis loop as long as enough number of MCS$/$K is used (e.g. $10^6$).
		
		\subsection{Temperature-Induced SCO Hysteresis}
		\begin{figure}[bp]
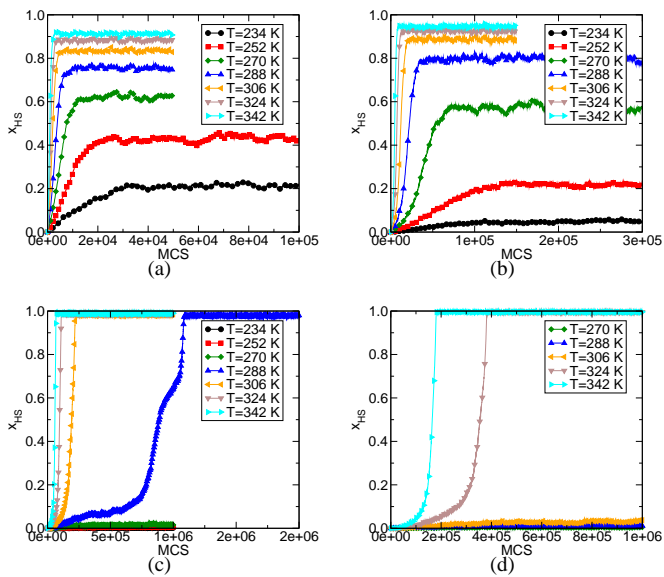

			\begin{tabular}{cc}
			\centering
			\subfigure[]{\includegraphics[width=0.23\textwidth]{conv-test-10.eps}}	&
			\subfigure[]{\includegraphics[width=0.23\textwidth]{conv-test-20.eps}}	\\
			\subfigure[]{\includegraphics[width=0.23\textwidth]{conv-test-40.eps}}	&
			\subfigure[]{\includegraphics[width=0.23\textwidth]{conv-test-50.eps}}
			\end{tabular}
			\caption{Convergence tests of $k_1=1.8\times{}10^4\,$K (a), $3.6\times{}10^4\,$K (b), $7.2\times{}10^4\,$K (c) and $9.0\times{}10^4\,$K (d). It is easy to see that more MCSs are required when (i) $k_1$ is larger, and (ii) temperatures are close to phase transition.}
			\label{fig:convergence test}
		\end{figure}		
		Based on the verification of convergence properties of the SAB model, we can now exploit its ability to simulate hysteresis phenomena. First we tackle the temperature-induced SCO under normal pressure. In order to estimate the appropriate sweeping rates, we depict HS fraction $x_{\mr{HS}}$ versus $\mr{MCS}$ for various values of $k_1$ at several representative temperatures (FIG. \ref{fig:convergence test}). When $k_1$ is small, e.g. (a) and (b), a gradual increase of stable $x_{\mr{HS}}$ is observed as temperature goes up, which indicates a gradual phase transition curve $x_{\mr{HS}}(T)$. On the other hand, when $k_1$ is large, e.g. (c) and (d), no intermediate values between $0$ and $1$ of stable $x_{\mr{HS}}$ are observed and abrupt phase transitions with hysteresis are expected.
						   
\begin{figure}[tp]
			\centering
			\includegraphics[width=0.40\textwidth]{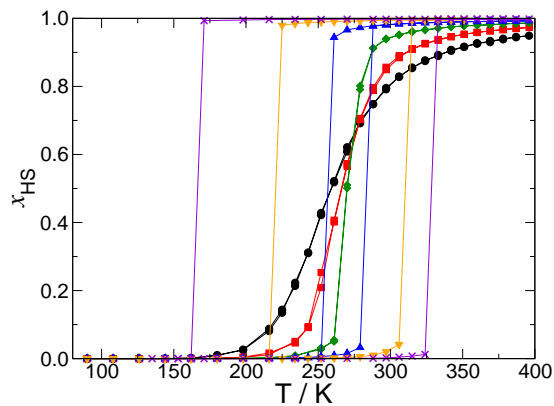}
			\caption{Phase transition curves $x_{\mr{HS}}(T)$ with $k_1=1.8\times{}10^4\,$K (black circle), $3.6\times{}10^4\,$K (red square), $5.4\times{}10^4\,$K (green diamond), $7.2\times{}10^4\,$K (blue triangle-up), $9.0\times{}10^4\,$K (orange triangle-down), and $1.1\times{}10^5\,$K (purple cross).}
			\label{fig:assemble_polySHO}
		\end{figure}

		After preparing all of these, we study the shapes of $x_{\mr{HS}}(T)$ curves with respect to different interaction strength $k$'s, and the results are depicted in FIG. \ref{fig:assemble_polySHO}. As we expected, $k_1=1.8\times{}10^4\,$K and $3.6\times{}10^4\,$K result in gradual changes, while $k_1\geq{}7.2\times{}10^4\,$K give abrupt phase transitions with hysteresis. The case of $k_1=5.4\times{}10^4\,$K seems to be somewhat critical, also abrupt but without loop. The change of $x_{\mr{HS}}(T)$ curves from being gradual to abrupt with increasing interactions follows the concept of ``generic sequence" raised by Y. Konishi.\cite{Konishi08} 

		It is worth pointing out that, in the simulations above we only employ the ratio (\ref{eq:ratio_k}) for $k_1$ and $k_2$ obtained from DFT computations rather than the absolute values. If the latter is used, i.e. $k_1=1.2\times{}10^6\,$K and $k_2=5.5\times{}10^5\,$K, a much wider loop is expected, even with HS metastable state in low temperature areas. A possible explanation to this inconsistency is the anisotropy (i.e. tetragonal rather than cubic symmetry) of the compound \ce{[Fe(pz)Pt(CN)4].2H2O}. In principle, introducing of more $k$'s to describe this anisotropy is likely to improve the situation. Nevertheless, as an example, the SAB model has already demonstrate its ability to link its key parameters with realistic quantities. 
		
		
		\subsection{Pressure Effects on SCO Hysteresis}
\begin{figure}[btp]
			\centering
			\includegraphics[width=0.40\textwidth]{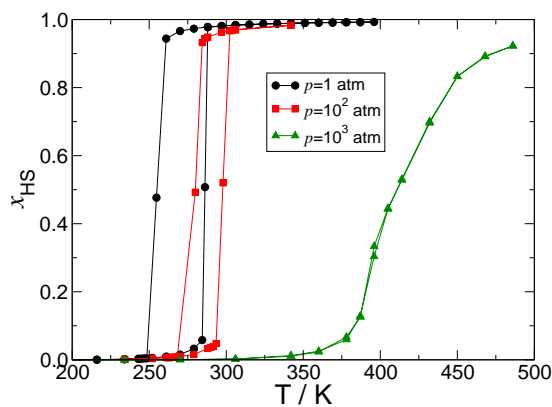}
			\caption{Simulating pressure effects on hysteresis loop with the SAB model for the case $k_1=7.2\times{}10^4\,$K.}
			\label{fig:pressure_effects}
		\end{figure}

		Since in the SAB model the pressure $p$ is an adjustable parameter, we can investigate the pressure effects on hysteresis loops. We choose the case $k_1=40$ in the above simulations and consider its behaviours when increasing $p$ from $1\,$atm to $10^3\,$atm. The results are depicted in FIG. \ref{fig:pressure_effects}. At first an increase of pressure shrinks the loop and shifts it to a higher $T_{1/2}$. By continuing increasing $p$, the loop moves to the right further and finally vanishes, turning into a gradual closed curve. These results are easy to understand since the LS state has a smaller volume and is stabilized under higher pressure, leading to a shrinking loop with higher $T_{1/2}$. 
		
	\section{Concluding Remarks}
		In summary, the SAB model is shown to be both free of the convergence problem and able to simulate the temperature-induced SCO phenomenon as well as pressure effects. All the parameters in the SAB model are demonstrated to be linked to experimental observables or \textit{ab initio} results, making it able to describe the realistic SCO process in principle.
		
		However, in this work only qualitative consistency is achieved with $k_1$ and $k_2$ predicted by DFT calculations. Apart from the anisotropy discussed above, the SAB model itself is still too simple. Although both the vibrational (up to the three-body level) and vibronic interactions have been introduced through stretching and bending oscillators, anharmonic contributions are absent. 
		
		Albeit on the one hand, the crossing terms in eqn (\ref{eq:expansion_H_1}) give an explicit expression for the vibronic interactions for the first time, which partially accounted for the spin-dependent elastic interaction terms in the Nasser-Boukheddaden (N-B) model. \cite{J.A.Nasser01,J.A.Nasser04,J.A.Nasser05,J.A.Nasser11,K.Boukheddaden07} On the other hand, since $k$'s optimized from simulations are smaller than those obtained from DFT calculations, we might conclude that a weak interaction would lead to a rather wide hysteresis loop too. This could somewhat explain the hysteresis behaviour of some molecular SCO compounds. Actually if those $k$'s (i.e. $k_1=2k_2=7.2\times{}10^4\,$K) are substituted into eqn (\ref{eq:modulus}), we can have an estimate for the bulk modulus $K\approx{}3\,$GPa and shear modulus $G\approx{}5\,$GPa, which correspond to relatively soft materials. 
				
		
		More importantly, noting that $\theta_0$ in (\ref{eq:polymer SHO model}) can assume values other than $\pi/2$, the SAB model is by nature applicable to describe materials with non-right-angle symmetries. Also, an extension to lattices of low dimensionality or other boundary conditions is straightforward. Thus finite-size effects and low dimensional effects can be exploited using the SAB model too.

\section*{Acknowledgement}
       	We would like to thank Mr. Zehua Chen in our research group. He provided us with many useful suggestions on our codes. We would also like to thank Prof. Tsung-Dao Lee and Mrs. Hui-Chun Chin, for their great generosity and encouragement on Chinese youth researchers. This project was supported by the Hui-Chun Chin and Tsung-Dao Lee Chinese Undergraduate Research Endowement (CURE), National Natural Science Foundation of China (Projects No. 20973009, 21373017), National Basic Research Program of China (2013CB933400) and Ministry of Education of China (20120001110063).

\appendix

	\section{Deriation of Eq. (\ref{eq:modulus}) and (\ref{eq:k1_and_k2})}					
		Here we show the derivation details for Eqs. (\ref{eq:modulus}) and (\ref{eq:k1_and_k2}). For the bulk modulus $K$, consider a homogeneous expansion of a simple cubic lattice which has volume $V_0$ at equilibrium and now $V$. The total electronic energy corresponding to this volume change is:
		\begin{equation}	\label{eq:EOS_V}
			\begin{split}
			E_{\mr{el}}(V)
				&=3N\times{}\frac{k_1}{2}
				\bigg[\bigg(\frac{V}{N}\bigg)^{1/3}-
				\bigg(\frac{V_0}{N}\bigg)^{1/3}\bigg]^{2}				\\
				&=\frac{3}{2}Nk_1\bigg(\frac{V_0}{N}\bigg)^{1/3}
				\bigg[\bigg(\frac{V}{V_0}\bigg)^{1/3}-1\bigg]^2			\\
				&=\frac{3}{2}Nk_1v_0^{2/3}(x-1)^2
			\end{split}
		\end{equation}
		where $v_0=V_0/N$ is the unit cell volume and $x=(V/V_0)^{1/3}$ is the stretching ratio in length. This is actually the EOS of volume change appeared in (\ref{eq:k1_and_k2}). Substitute (\ref{eq:EOS_V}) into the definition of bulk modulus $K=V\md^2E/\md{}V^2$, we have
		\begin{equation}
			K=\frac{1}{3}\bigg(\frac{N}{V}\bigg)^{1/3}
				\bigg[2\bigg(\frac{V_0}{V}\bigg)^{1/3}-1\bigg]k_1.
		\end{equation}
		At equilibrium, $V=V_0$ and hence $K=k_1/(3v_0^{1/3})$ holds.
		
		For the shear modulus $G$, consider a small displacement of $\gamma$ from equilibrium position $\gamma_0=\pi/2$. The total electronic energy is
		\begin{equation}	\label{eq:EOS_gamma}
			E_{\mr{el}}(\gamma)=4N\times{}\frac{k_2}{2}
				(\gamma-\gamma_0)^2,
		\end{equation}
		which is the EOS of $\gamma$ in (\ref{eq:k1_and_k2}). According to the definition of shear modulus, we have
		\begin{equation}
			G = \frac{1}{\gamma-\gamma_0}\frac{\md{}E}{\md{}\gamma}=4Nk_2,
		\end{equation}
		which completes the derivations for (\ref{eq:modulus}).




\footnotesize{
\bibliography{refs-latest} 
\bibliographystyle{rsc} 
}

\end{document}